# The absence of Bohr – van Leeuwen paradox in classical statistical ensembles of moving charges, in finite phase volume.

A. N. Pechenkov

Institute of Metal Physics, Ural Division of the Rus. Acad. Sci., 18,
S.Kovalevskaya str., GSP – 170, 620041,Ekaterinburg, Russia.
pechenkov@imp.uran.ru

The Bohr – van Leeuwen theorem [1-6] consists in follow paradox: classical statistical ensambles of moving charges in external static magnetic field can't have the induced orbital magnetic moment. I.e., the diamagnetism is not possible.
In that paper will be shown, the theorem take place only in statistical ensembles in infinite phase volume. For the statistical ensembles in finite phase volume we have usual diamagnetic orbital moment.

Keywords:

First of all, let us consider the Bohr – van Leeuwen theorem, follow by [4]. The resulting orbital magnetic moment of moving charges is:

$$\mathbf{p}_{mag} = \frac{1}{2c}\sum_i q_i [\mathbf{r}_i, \mathbf{v}_i] \qquad (1)$$

here are: c – vacuum light velocity; q – value of a charge; **r** – coordinate vector of a charge; **v** – velocity vector of a charge.

The magnetic moment we can to formulate as statistical average value in thermal equilibrium ensemble of point charges in external static magnetic field, without own spin magnetic moments.

As it is predicted by Maxwell – Boltzmann statistic, the average value of any function is:

$$\overline{f} = \frac{\int_{-\infty}^{\infty} f(\mathbf{r},\mathbf{p}) \exp(-\frac{\mathsf{H}(\mathbf{r},\mathbf{p})}{kT}) d\mathbf{r} d\mathbf{p}}{Z} \qquad (2)$$

$$Z = \int_{-\infty}^{\infty} \exp(-\frac{\mathsf{H}(\mathbf{r},\mathbf{p})}{kT}) d\mathbf{r} d\mathbf{p} \qquad (3)$$

here are: **p** – mechanical moment of a charge; k – Boltzmann constant; T – temperature of the ensemble; $\mathsf{H}$ – Hamiltonian of the ensemble.

Emphasize here the infinite integral limits in phase space! Let the charges have a weak interactions one with another. Then the charges are statistical independent ones, and we can to consider only one charge. We will consider the motion of a charge in the plane Z=0, with external magnetic field along Z – axis.
Then Hamiltonian of a charge is:

$$\mathsf{H} = \frac{1}{2m}\left[\left(p_x - \frac{q}{c}A_x\right)^2 + \left(p_y - \frac{q}{c}A_y\right)^2\right] \qquad (4)$$

here: **A** is magnetic vector potential, m – mass of a charge.
We have further:

$$\mathbf{B} = rot\mathbf{A} = (0,0,B) \Rightarrow \mathbf{A} = \frac{B}{2}(-y,x,0) \tag{5}$$

Therefore:

$$\mathsf{H} = \frac{1}{2m}\left[\left(p_x + \frac{q}{2c}By\right)^2 + \left(p_y - \frac{q}{2c}Bx\right)^2\right] \tag{6}$$

Orbital magnetic moment of moving charges (1) is:

$$\mathbf{p}_{mag} = (0,0,\frac{q}{2c}(xV_y - yV_x)) \tag{7}$$

Velocity components of a charge can be found from the Hamilton equation:

$$\mathbf{V} = \frac{\partial \mathsf{H}}{\partial \mathbf{p}} \Rightarrow V_x = \frac{1}{m}\left(p_x + \frac{q}{2c}By\right); V_y = \frac{1}{m}\left(p_y - \frac{q}{2c}Bx\right); V_z = 0; \tag{8}$$

Then we have:

$$\begin{aligned}\mathbf{p}_{mag} &= (0,0,\frac{q}{2c}(x\frac{1}{m}\left(p_y - \frac{q}{2c}Bx\right) - y\frac{1}{m}\left(p_x + \frac{q}{2c}By\right))) = \\ &= (0,0,\frac{q}{2mc}(x\left(p_y - \frac{q}{2c}Bx\right) - y\left(p_x + \frac{q}{2c}By\right)))\end{aligned} \tag{9}$$

We ca to see now the follow expression for the $p_{mag\ z}$:

$$p_{mag\ z} = -\frac{\partial \mathsf{H}}{\partial B} \tag{10}$$

Or, in the vector view, we have:

$$\mathbf{p}_{mag} = -\frac{\partial \mathsf{H}}{\partial \mathbf{B}} \tag{11}$$

Now, the mean value of the orbital magnetic moment can be written as:

$$\overline{p_{mag\ z}} = -\frac{\int_{-\infty}^{\infty} \frac{\partial \mathsf{H}}{\partial B} \exp(-\frac{\mathsf{H}(\mathbf{r},\mathbf{p})}{kT})d\mathbf{r}d\mathbf{p}}{Z} = kT\frac{\partial \ln Z}{\partial B} \tag{12}$$

Bohr – van Leeuwen theorem tell us:

$$\overline{p_{mag\ z}} = kT\frac{\partial \ln Z}{\partial B} = 0 \tag{13}$$

i.e., the mean value of the orbital magnetic moment of a charge is zero.
To proof (13), we must only to change the independent variables in the integral $Z$ :

$$g_x = p_x - \frac{q}{c}A_x; \quad g_x = p_y - \frac{q}{c}A_y; \tag{14}$$

It is light to see, that after the substitution, we eliminate the vector potential **A** from the integral Z. Therefore, the derivative in (13) will be zero. I.e., the Bohr – van Leeuwen theorem is proofed.

Therefore, we have the absence of the diamagnetism. The paradox is the consequence of the infinite limits in integral Z, because magnetic vector potential **A** can be included in the limits after the substitution (14), in the case of finite phase volume of moving charge. Let us consider below the last case.

Let the upper modulus of every coordinate is L, and the upper modulus of every component of the mechanical moment is P. Then we have:

$$Z = \int_{-L}^{L} dxdy \left[ \int_{-P}^{P} dp_x dp_y \exp\left( -\frac{1}{2mkT}\left[ \left(p_x + \frac{q}{2c}By\right)^2 + \left(p_y - \frac{q}{2c}Bx\right)^2 \right] \right) \right] \tag{15}$$

After the substitution (14), we have from (15):

$$Z = \int_{-L}^{L} dxdy \left[ \int_{-P-\frac{q}{2c}Bx}^{P-\frac{q}{2c}Bx} \exp\left(-\frac{g_y^2}{2mkT}\right) dg_y \int_{-P+\frac{q}{2c}By}^{P+\frac{q}{2c}By} \exp\left(-\frac{g_x^2}{2mkT}\right) dg_x \right] \tag{16}$$

Now, using the Leibniz rule, we have:

$$\frac{\partial Z}{\partial B} = \int_{-L}^{L} dxdy \frac{\partial}{\partial B} \int_{P1}^{P2}\int_{P3}^{P4} F(g_x, g_y) dg_x dg_y = \int_{-L}^{L} dxdy \left[ \begin{array}{l} \int_{P1}^{P2}\int_{P3}^{P4} \frac{\partial F(g_x,g_y)}{\partial B} dg_x dg_y + \frac{\partial P_2}{\partial B}\int_{P3}^{P4} F(P_2, g_y) dg_y - \\ -\frac{\partial P_1}{\partial B}\int_{P3}^{P4} F(P_1, g_y) dg_y + \frac{\partial P_4}{\partial B}\int_{P1}^{P2} F(g_x, P_4) dg_x - \\ -\frac{\partial P_3}{\partial B}\int_{P1}^{P2} F(g_x, P_3) dg_x \end{array} \right]$$

In our case: $\dfrac{\partial F(g_x, g_y)}{\partial B} = 0$. Then:

$$\frac{\partial Z}{\partial B} = \frac{q}{2c}\int_{-L}^{L} dxdy \left[ \begin{array}{l} y\left[\exp\left(-\frac{(P+\frac{q}{2c}By)^2}{2mkT}\right) - \exp\left(-\frac{(-P+\frac{q}{2c}By)^2}{2mkT}\right)\right] \int_{-P-\frac{q}{2c}Bx}^{P-\frac{q}{2c}Bx} \exp\left(-\frac{g_y^2}{2mkT}\right) dg_y + \\ +x\left[\exp\left(-\frac{(-P-\frac{q}{2c}Bx)^2}{2mkT}\right) - \exp\left(-\frac{(P-\frac{q}{2c}Bx)^2}{2mkT}\right)\right] \int_{-P+\frac{q}{2c}By}^{P+\frac{q}{2c}By} \exp\left(-\frac{g_x^2}{2mkT}\right) dg_x \end{array} \right]$$

$$\tag{17}$$

Let the parameters : L, P, B have such values, that $g_x \sim g_y \sim 0$. Then we can to approximate: $\exp(-\frac{g^2}{2mkT}) \cong 1 - \frac{g^2}{2mkT}$. Then we have from (16,17):

$$\frac{\partial Z}{\partial B} = \frac{q}{2c}\int_{-L}^{L} dxdy \left[ y\left[-\frac{qByP}{mckT}\right] \int_{-P-\frac{q}{2c}Bx}^{P-\frac{q}{2c}Bx} \exp\left(-\frac{g_y^2}{2mkT}\right)dg_y + x\left[-\frac{qBxP}{mckT}\right] \int_{-P+\frac{q}{2c}By}^{P+\frac{q}{2c}By} \exp\left(-\frac{g_x^2}{2mkT}\right)dg_x \right] =$$

$$= \frac{q}{2c}\int_{-L}^{L} dxdy \left[ y\left[-\frac{qByP}{mckT}\right]2P + x\left[-\frac{qBxP}{mckT}\right]2P \right] = -\frac{1}{mkT}\left(\frac{q}{c}\right)^2 P^2 B \int_{-L}^{L} dxdy[y^2 + x^2] =$$

$$= -\frac{8}{3}L^4 \frac{q^2}{mc^2 kT} P^2 B$$

(18)

In (18) we used the linear approximation of the integrals. Also from (16) follows:

$$Z = 16(LP)^2 \qquad (19)$$

Therefore, from (12) follows mean value of the orbital magnetic moment of a charge, in external magnetic field:

$$p_{mag\ z} = -\frac{1}{6}\frac{(qL)^2}{mc^2}B \qquad (20)$$

It is well known formula for the diamagnetism [7].

Therefore, the Bohr – van Leeuwen theorem is valid only for the classical statistical ensambles in infinite phase volume.

Note, the Bohr – van Leeuwen theorem is formulated for a moving charge without radiation reaction. I.e. the radiation reaction is not included into Hamiltonian. Such charge can to move in external magnetic field infinite long time. If we will include the radiation reaction into Hamiltonian, than a charge will loses it's kinetic energy, and the diamagnetism will exists a finite long time. But it is not equilibrium process. The Bohr – van Leeuwen theorem is formulated for an equilibrium process.